\documentclass{article} 
\usepackage{iclr2020_conference,times}

\usepackage{amsmath,amsfonts,bm}

\def\eqref#1{equation~\ref{#1}}

\def\1{\bm{1}}

\DeclareMathAlphabet{\mathsfit}{\encodingdefault}{\sfdefault}{m}{sl}
\SetMathAlphabet{\mathsfit}{bold}{\encodingdefault}{\sfdefault}{bx}{n}

\usepackage{hyperref}
\usepackage{url}
\usepackage{enumitem}
\usepackage{comment}
\usepackage{graphicx}

\title{Participatory Problem Formulation for Fairer Machine Learning Through Community Based System Dynamics}

\author{Donald Martin, Jr.\\
Google \\
\texttt{dxm@google.com} \\
\And
Vinodkumar Prabhakaran \\
Google \\
\texttt{vinodkpg@google.com} \\
\And
Jill Kuhlberg \\
System Stars \\
\texttt{jill@systemstars.co} \\
\And
Andrew Smart \\
Google \\
\texttt{andrewsmart@google.com} \\
\And
William S. Isaac \\
DeepMind \\
\texttt{williamis@google.com} 
}

\iclrfinalcopy 
\begin{document}

\maketitle

\begin{abstract}
Recent research on algorithmic fairness has highlighted that the problem formulation phase of ML system development can be a key source of bias that has significant downstream impacts on ML system fairness outcomes. However, very little attention has been paid to methods for improving the fairness efficacy of this critical phase of ML system development. Current practice neither accounts for the dynamic complexity of high-stakes domains nor incorporates the perspectives of vulnerable stakeholders. 
In this paper we introduce community based system dynamics (CBSD) as an approach to enable the participation of  typically excluded stakeholders in the problem formulation phase of the ML system development process and facilitate the deep problem understanding required to mitigate bias during this crucial stage.
\end{abstract}

\section{Introduction}

Problem formulation is a crucial first step in any machine learning (ML) based interventions that have the potential of impacting the real lives of people;
a step that involves 
determining the strategic goals driving the interventions and 
translating those strategic goals into tractable machine learning problems \citep{barocas2017big,passi2019problem}. 
The decisions made during this step can have profound impact in shaping the core aspects of those interventions, including how they impact different communities in society. Recent studies have demonstrated many instances where ML-aided interventions in high-stakes domains such as health-care risk-assessment \citep{obermeyer2019dissecting}, criminal justice \citep{chouldechova2017fair} and online content moderation \citep{sap-etal-2019-risk} resulted in unintended unfair outcomes that further disadvantaged already marginalized communities. 

Researchers have pointed out two major pitfalls in the problem formulation step that contributes to such undesirable outcomes. 
First, the problem formulation step necessitates developing a model of the problem domain that accommodates the constraints of leveraging existing ML tools (often the pre-chosen intervention method), most of which operate on (i.e., classifies or regresses over) data that are static snap-shots of the problem domain, and consequently often ignores the non-linear and dynamically complex nature of society that involves feedback loops and time-delays between actions and impacts \citep{ensign2017runaway, mortiz2018}.
Second, the stakeholders that are involved in problem formulation --- e.g., product managers, business analysts, computer scientists, ML practitioners --- often lack the lived experiences required to comprehensively approximate and account for the various peripheral stakeholders their interventions will impact, especially the communities that are the most vulnerable to unfair outcomes \citep{eubanks2018automating, campolo2017ai}.

For instance, let us consider the recent study \citep{obermeyer2019dissecting} which discovered that a prediction algorithm broadly used by health-care risk-assessment tools exhibited racial bias against African-Americans. 
The strategic goal of those tools was to improve the care of patients with complex health needs while reducing overall costs, by targeting high-risk patients (i.e., those with complex health needs) with special programs and resources. The goal itself implies an interventionist \citep{ben2018causation} approach that relies on the causal inference that \textit{special programs for high-risk patients will lead to or cause lower overall, system-wide healthcare costs}.
During the problem formulation phase, this strategic goal was reduced to identifying the patients who had the highest health care costs; essentially using costs/spending as a proxy for needs. This reduction relied on the additional human causal inference that \textit{patients --- regardless of population or background --- with more complex health needs would have spent more on health care in the past and that no other factors were causally relevant to their health care spending}. However, this inference proved incorrect as it failed to consider a multitude of confounding factors and the dynamically complex ways they impact health care spending.  Specifically, the historic disparities in health care access (among other things) that African American individuals face in the US healthcare system means that they end up spending less on health care. Consequently, the health risk-assessment algorithm tended to mistakenly infer African American individuals to not be high-risk patients, regardless of the complexity of their illnesses, further denying them access to special programs and resources. 

One of the core mistakes made in the above scenario is in the problem formulation step itself --- using health-care costs (spending) as a proxy variable for health-care need. The causal theories that guided this process emerge from an opaque and iterative process among key stakeholders (e.g. product managers, executives, business analysts) that we collectively refer to as "causal reasoners" \citep{pearl2018book}.  Psychological research has shown that human causal inference is based on a priori intuitive theories about the causal structure of the problem to be intervened on \citep{tenenbaum2003theory,pearl2018book}. These \textit{causal theories} are the result of the cumulative lived experiences of the individual causal reasoner and reflect views of reality filtered through their world views and biases. If the problem formulation step had facilitated the equitable participation of
African American community members who have lived experience within the US healthcare system, it is likely this undesired outcome could have been averted.

While researchers have recognized the importance of problem formulation in ensuring fair and ethical machine learning interventions in society, the process that guides this step still remains ad-hoc, informal, and fueled by intuition \citep{barocas2016big,barocas2017big}. Such reliance on the implicit causal theories of causal reasoners,  who may lack the lived experiences required to comprehensively approximate the causal model of the problem domain upon which to base inferences, will continue to result in undesirable outcomes. Hence, the problem formulation step, especially in high-stakes situations, should have at its core a formal approach to developing causal models of the socio-technical problem domains being intervened upon. Such an approach should incorporate two key attributes: (1) ability to contend with the delayed impacts and feedback loops that characterize the dynamically complex nature of high-stakes contexts, and (2) optimized for making causal inferences explicit and for iterative learning of causal structures in partnership with peripheral stakeholders including policy makers and social groups most vulnerable to unfair outcomes.

\section{Community Based System Dynamics}

In this paper we introduce \textit{Community Based System Dynamics (CBSD)} as an approach to engage multiple and diverse stakeholder groups in problem formulation to design fairer ML-based interventions. CBSD is a participatory method for involving communities in the process of developing a shared understanding of complex systems from the feedback perspective. It relies on visual tools and simulation to support groups in the co-development of explicit and transparent causal theories \citep{hovmand2014community}.  CBSD's explicit goal to build capacity among stakeholders to derive deeper system insights together through their participation sets it apart from other approaches where stakeholders are viewed as informants. It has been used to engage and center the perspectives of marginalized and vulnerable communities in the development of more effective interventions in ecology, public health, and social work \citep{stave2015system, trani2016community, escobedo2019community}.

Unlike other causal modeling techniques such as Structural Causal Models (SCMs) and Causal Bayesian Networks (CBNs) that have been recently proposed to model causality in ML problem formulation \citep{madras2019fairness,chiappa2018causal}, CBSD is founded upon a system dynamics (SD) \citep{hovmand2014community} approach, which takes a characteristically feedback approach to modeling dynamically complex problems \citep{sterman2010business, richardson2011reflections}. In addition, while CBNs and SCMs are independent causal modeling tools,
combining causal modeling tools with formal practices for collaborative and iterative causal modeling is inherent to both SD and CBSD.

\begin{figure*}
  \centering
    \includegraphics[width=0.65\textwidth]{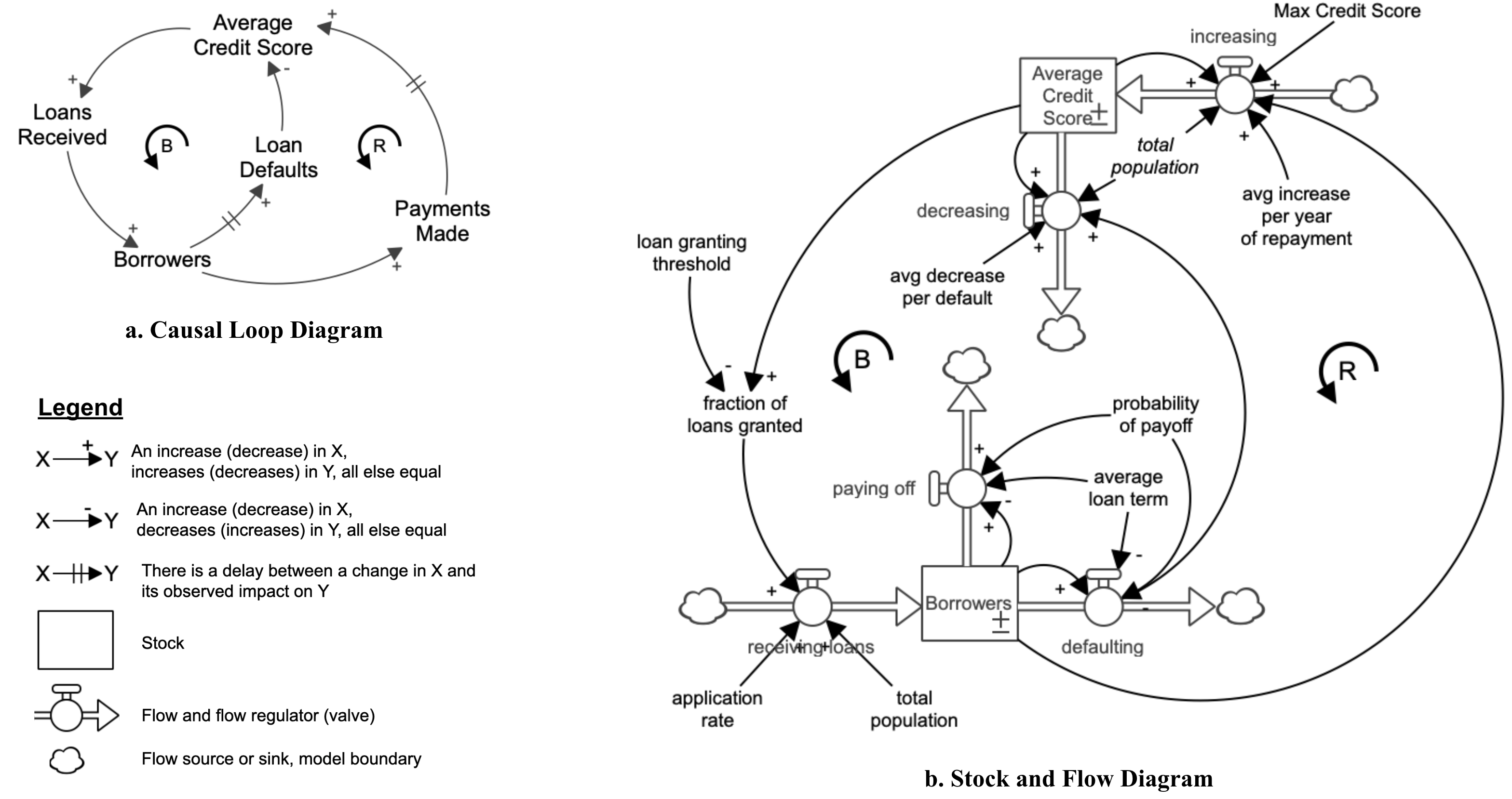}
    \caption{Examples of system dynamics causal loop diagrams and stock and flow diagrams.\label{fig_sd_loans_moritz}}
\end{figure*}


\subsection{Causal Modeling using System Dynamics}

System Dynamics (SD) is defined as \textit{the process of using both informal maps/diagrams and formal models with computer simulation to uncover and understand the dynamics of complex problems from a feedback perspective} \citep{richardson2011reflections}. It is this emphasis on feedback --- reinforcing and balancing processes that unfold over time --- that distinguishes SD from other causal modeling approaches, and makes it apt for the dynamically complex nature of high-stakes problems at the center of risk-prediction systems. To uncover and understand feedback processes, SD has developed a series of tools that vary in degree of formalism and are designed to provide insight into different aspects of the complex problems they model.
Many of these tools are graphical in nature, requiring modelers to make their causal theories explicit, thereby fostering transparency \citep{lane2008emergence}.

One of the most commonly used visual tools in SD is the causal loop diagram (CLD). The main purpose of the CLD is to show the feedback processes in a system (understood as the set of posited causal structures related to the phenomenon of interest) using a directed graph. CLDs are often used to quickly elicit hypothesized causal relations between variables in a problem space and/or communicate the main feedback loops in a more detailed computer simulation model. 

An example of a CLD is shown in Figure~\ref{fig_sd_loans_moritz}a, which offers a simplified representation of a credit score based lending system. Such systems resemble the health-risk system described earlier in that they utilize risk prediction algorithms to intervene on high-stakes problem domains with vulnerable stakeholders.  The arrows in CLDs represent hypothesized causal links between variables, with the arrowheads and polarity indicating the direction and the nature of influence. Positive polarity represents relationships where an increase (decrease) in one variable triggers an increase (decrease) in the other, all else equal. Negative polarity is used to depict relationships where an increase (decrease) in one variable triggers a decrease (increase) in the other, all else equal.  In the example in Figure~\ref{fig_sd_loans_moritz}a, the relationship between \emph{Payments Made} and \emph{Average Credit Score} is assumed to be of positive polarity since making payments towards debt generally builds credit, \emph{ceteris paribus}, whereas the link between \emph{Loan Defaults} and \emph{Average Credit Score} is negative, since defaulting generally results in score reductions. Any increase (decrease) in the average credit score of a group leads to a corresponding increase (decrease) in the number of loans received by that group, which in-turn  increases (decreases) its borrower pool. 

A more formal treatment of the causal structures, including the concept of delays and their impact on the system is offered by stock and flow diagrams, perhaps the most commonly used tool in system dynamics.  In addition to representing relationships between variables and feedback loops, stock and flow diagrams require explicit definitions of variables that vary, and the precise ways that they accumulate or are depleted over time.  In these diagrams, variables that accumulate are called \textit{stocks} and are drawn as rectangles, and the processes that add to or drain them are called \textit{flows} (inflows and outflows) and are depicted as double-lined/thick arrows or ``pipes'' with valves. The ``clouds'' are the sources and sinks of the flows, and are assumed to have infinite capacity over the time horizon of the model.  These clouds show the model's assumed boundary --- once information or material passes through the flows into a cloud, it ceases to impact the system.


Figure~\ref{fig_sd_loans_moritz}b shows a stock and flow representation of the lending system represented in the CLD (Figure~\ref{fig_sd_loans_moritz}a), in which \emph{Borrowers} and  \emph{Average Credit Score} of the population are now represented as stocks, and are thus assumed to accumulate value over time. The number of borrowers (units = people) accumulates the inflow of people \textit{receiving loans} per year and is depleted by the outflows of people \emph{paying off} the loan completely and \emph{defaulting} on loans per year. In this context the cloud before \emph{receiving loans} indicates that there is an endless source of individuals who could apply for loans. In turn, those leaving the system, by defaulting or paying off,  are assumed to not affect the system in any meaningful way, and are thus represented as clouds at the ends of the outflows.

Stock and flow diagrams serve a qualitative purpose, but they are also the critical bridge to simulation and the quantitative aspect of SD.  Visualizing behavior over time is extremely difficult with static diagrams. Simulation enables the visualization and validation of dynamic hypotheses of causal structures upon which interventions are ultimately based.  Additionally, simulation is a critical step for gaining a deep understanding of dynamic causal structures and enables "en silico" intervention experimentation.



\subsection{Participatory Modeling}

SD has a rich history of participatory modeling that involves stakeholders in the model building process to foster collaboration and learning \citep{kiraly2019dynamics}. Community based system dynamics \citep{hovmand2014community} is a particular SD practice approach that engages stakeholders who are embedded in the system of interest to conceptualize the problem, identify the related issues and prioritize interventions based on model supported insights. More than just involving participants in the modeling process to elicit information, CBSD has the explicit goal of building capabilities within communities to use SD on their own, distinguishing it from other participatory approaches in SD. Building capabilities enables stakeholders to more accurately represent their causal theories in the models, which is especially critical when the stakeholders are from marginalized communities that are not represented in the modeling process. In this view, individual and community perspectives on the structures that underlie everyday experiences are valued as valid and necessary sources of data, and community perspectives on the analysis and interpretation of models is essential for realizing the value of the approach. 

Best practices for engaging stakeholders in the process of reflecting problem causal structure using CLDs and stock and flow diagrams and refining simulation models are documented \citep{hovmand2014community, hovmand2014group}. These activities can be adapted for diverse contexts and support the development of capabilities for collaborative causal inference development.  Overall, CBSD has been shown to be useful in a broad range of problem domains such as maternal and child health \citep{munar2015scaling}, identifying food system vulnerabilities \citep{stave2015system}, mental health interventions \citep{trani2016community} and alcohol abuse \citep{apostolopoulos2018moving}, to name a few. 
In the domain of ML (un)fairness, the use of CBSD can help center the voices and lived experiences of those marginalized communities that are potentially impacted by ML-based products. If the goal is to design fairer ML-based tools and products that do not harm peripheral stakeholders, it is imperative to not only model the long-term dynamics created by those products, but to also incorporate the perspectives of those stakeholders in defining what fairness means in a particular domain or context.

\section{Discussion and Conclusion}

In this paper we highlight the causal inferences of key stakeholders as the central point of risk in the currently adhoc and informal problem formulation process.  
When the process that generates those causal inferences is opaque and insular real harms can result.  We introduced CBSD as a mature candidate to foster the formalization  of the problem formulation process in a manner that considers the dynamically complex nature of high-stakes contexts and enables the diversification of the sources of causal theories upon which human causal inferences are ultimately based.  A key advantage of an SD-based approach is that it draws heavily on visual diagramming conventions which emphasize transparency and facilitate the engagement of diverse stakeholders to add, revise and critique causal theories. 
A long lineage of participatory approaches within SD, including CBSD and group model building, provide evidence of success in developing and using system dynamics models in diverse contexts, and serve as resources for groups interested in developing SD capabilities in their communities/contexts. 
Moreover, a strength SD shares with other causal modeling approaches, including Bayesian networks, 
is the correspondence between its visualizations and their underlying mathematical representations, which allows stakeholders to do more than visualize, but continue to develop deep insights about important data to collect and consider, as well as evaluate the impact of products and decisions through simulation. 

\subsubsection*{Acknowledgments}
We would like to thank 
Emily Denton,
Ben Hutchinson,
Sean Legassick,
Silvia Chiappa,
Matt Botvinick,
Reena Jana, 
Dierdre Mulligan, and
Deborah Raji
for their valuable feedback on this paper.
\bibliography{problemform4}

\begin{thebibliography}{26}
\providecommand{\natexlab}[1]{#1}
\providecommand{\url}[1]{\texttt{#1}}
\expandafter\ifx\csname urlstyle\endcsname\relax
  \providecommand{\doi}[1]{doi: #1}\else
  \providecommand{\doi}{doi: \begingroup \urlstyle{rm}\Url}\fi

\bibitem[Apostolopoulos et~al.(2018)Apostolopoulos, Lemke, Barry, and
  Lich]{apostolopoulos2018moving}
Yorghos Apostolopoulos, Michael~K Lemke, Adam~E Barry, and Kristen~Hassmiller
  Lich.
\newblock Moving alcohol prevention research forward-part ii: new directions
  grounded in community-based system dynamics modeling.
\newblock \emph{Addiction}, 113\penalty0 (2):\penalty0 363--371, 2018.

\bibitem[Barocas \& Selbst(2016)Barocas and Selbst]{barocas2016big}
Solon Barocas and Andrew~D Selbst.
\newblock Big data's disparate impact.
\newblock \emph{Calif. L. Rev.}, 104:\penalty0 671, 2016.

\bibitem[Barocas et~al.(2017)Barocas, Bradley, Honavar, and
  Provost]{barocas2017big}
Solon Barocas, Elizabeth Bradley, Vasant Honavar, and Foster Provost.
\newblock Big data, data science, and civil rights.
\newblock \emph{arXiv preprint arXiv:1706.03102}, 2017.

\bibitem[Ben-Menahem(2018)]{ben2018causation}
Yemima Ben-Menahem.
\newblock \emph{Causation in Science}.
\newblock Princeton University Press, 2018.

\bibitem[Campolo et~al.(2017)Campolo, Sanfilippo, Whittaker, and
  Crawford]{campolo2017ai}
Alex Campolo, Madelyn Sanfilippo, Meredith Whittaker, and Kate Crawford.
\newblock Ai now 2017 report.
\newblock \emph{AI Now Institute at New York University}, 2017.

\bibitem[Chiappa \& Isaac(2018)Chiappa and Isaac]{chiappa2018causal}
Silvia Chiappa and William~S Isaac.
\newblock A causal bayesian networks viewpoint on fairness.
\newblock In \emph{IFIP International Summer School on Privacy and Identity
  Management}, pp.\  3--20. Springer, 2018.

\bibitem[Chouldechova(2017)]{chouldechova2017fair}
Alexandra Chouldechova.
\newblock Fair prediction with disparate impact: A study of bias in recidivism
  prediction instruments.
\newblock \emph{Big data}, 5\penalty0 (2):\penalty0 153--163, 2017.

\bibitem[Ensign et~al.(2017)Ensign, Friedler, Neville, Scheidegger, and
  Venkatasubramanian]{ensign2017runaway}
Danielle Ensign, Sorelle~A Friedler, Scott Neville, Carlos Scheidegger, and
  Suresh Venkatasubramanian.
\newblock Runaway feedback loops in predictive policing.
\newblock \emph{arXiv preprint arXiv:1706.09847}, 2017.

\bibitem[Escobedo et~al.(2019)Escobedo, Gonzalez, Kuhlberg, Calanche,
  Baezconde-Garbanati, Contreras, and Bluthenthal]{escobedo2019community}
Patricia Escobedo, Karina~Dominguez Gonzalez, Jill Kuhlberg, Maria~‘Lou’
  Calanche, Lourdes Baezconde-Garbanati, Robert Contreras, and Ricky
  Bluthenthal.
\newblock Community needs assessment among latino families in an urban public
  housing development.
\newblock \emph{Hispanic Journal of Behavioral Sciences}, 41\penalty0
  (3):\penalty0 344--362, 2019.

\bibitem[Eubanks(2018)]{eubanks2018automating}
Virginia Eubanks.
\newblock \emph{Automating inequality: How high-tech tools profile, police, and
  punish the poor}.
\newblock St. Martin's Press, 2018.

\bibitem[Hovmand(2014{\natexlab{a}})]{hovmand2014community}
Peter~S Hovmand.
\newblock \emph{Community Based System Dynamics}.
\newblock Springer, 2014{\natexlab{a}}.

\bibitem[Hovmand(2014{\natexlab{b}})]{hovmand2014group}
Peter~S Hovmand.
\newblock Group model building and community-based system dynamics process.
\newblock In \emph{Community Based System Dynamics}, pp.\  17--30. Springer,
  2014{\natexlab{b}}.

\bibitem[Kir{\'a}ly \& Miskolczi(2019)Kir{\'a}ly and
  Miskolczi]{kiraly2019dynamics}
G{\'a}bor Kir{\'a}ly and P{\'e}ter Miskolczi.
\newblock Dynamics of participation: System dynamics and participation--an
  empirical review.
\newblock \emph{Systems Research and Behavioral Science}, 36\penalty0
  (2):\penalty0 199--210, 2019.

\bibitem[Lane(2008)]{lane2008emergence}
David~C Lane.
\newblock The emergence and use of diagramming in system dynamics: a critical
  account.
\newblock \emph{Systems Research and Behavioral Science: The Official Journal
  of the International Federation for Systems Research}, 25\penalty0
  (1):\penalty0 3--23, 2008.

\bibitem[Liu et~al.(2018)Liu, Dean, Rolf, Simchowitz, and Hardt]{mortiz2018}
Lydia~T. Liu, Sarah Dean, Esther Rolf, Max Simchowitz, and Moritz Hardt.
\newblock Delayed impact of fair machine learning.
\newblock \emph{CoRR}, abs/1803.04383, 2018.
\newblock URL \url{http://arxiv.org/abs/1803.04383}.

\bibitem[Madras et~al.(2019)Madras, Creager, Pitassi, and
  Zemel]{madras2019fairness}
David Madras, Elliot Creager, Toniann Pitassi, and Richard Zemel.
\newblock Fairness through causal awareness: Learning causal latent-variable
  models for biased data.
\newblock In \emph{Proceedings of the Conference on Fairness, Accountability,
  and Transparency}, pp.\  349--358. ACM, 2019.

\bibitem[Munar et~al.(2015)Munar, Hovmand, Fleming, and
  Darmstadt]{munar2015scaling}
Wolfgang Munar, Peter~S Hovmand, Carrie Fleming, and Gary~L Darmstadt.
\newblock Scaling-up impact in perinatology through systems science: Bridging
  the collaboration and translational divides in cross-disciplinary research
  and public policy.
\newblock In \emph{Seminars in perinatology}, volume~39, pp.\  416--423.
  Elsevier, 2015.

\bibitem[Obermeyer et~al.(2019)Obermeyer, Powers, Vogeli, and
  Mullainathan]{obermeyer2019dissecting}
Ziad Obermeyer, Brian Powers, Christine Vogeli, and Sendhil Mullainathan.
\newblock Dissecting racial bias in an algorithm used to manage the health of
  populations.
\newblock \emph{Science}, 366\penalty0 (6464):\penalty0 447--453, 2019.

\bibitem[Passi \& Barocas(2019)Passi and Barocas]{passi2019problem}
Samir Passi and Solon Barocas.
\newblock Problem formulation and fairness.
\newblock In \emph{Proceedings of the Conference on Fairness, Accountability,
  and Transparency}, pp.\  39--48. ACM, 2019.

\bibitem[Pearl \& Mackenzie(2018)Pearl and Mackenzie]{pearl2018book}
Judea Pearl and Dana Mackenzie.
\newblock \emph{The book of why: the new science of cause and effect}.
\newblock Basic Books, 2018.

\bibitem[Richardson(2011)]{richardson2011reflections}
George~P Richardson.
\newblock Reflections on the foundations of system dynamics.
\newblock \emph{System Dynamics Review}, 27\penalty0 (3):\penalty0 219--243,
  2011.

\bibitem[Sap et~al.(2019)Sap, Card, Gabriel, Choi, and
  Smith]{sap-etal-2019-risk}
Maarten Sap, Dallas Card, Saadia Gabriel, Yejin Choi, and Noah~A. Smith.
\newblock The risk of racial bias in hate speech detection.
\newblock In \emph{Proceedings of the 57th Annual Meeting of the Association
  for Computational Linguistics}, pp.\  1668--1678, Florence, Italy, July 2019.
  Association for Computational Linguistics.
\newblock \doi{10.18653/v1/P19-1163}.
\newblock URL \url{https://www.aclweb.org/anthology/P19-1163}.

\bibitem[Stave \& Kopainsky(2015)Stave and Kopainsky]{stave2015system}
Krystyna~A Stave and Birgit Kopainsky.
\newblock A system dynamics approach for examining mechanisms and pathways of
  food supply vulnerability.
\newblock \emph{Journal of Environmental Studies and Sciences}, 5\penalty0
  (3):\penalty0 321--336, 2015.

\bibitem[Sterman(2010)]{sterman2010business}
John Sterman.
\newblock \emph{Business dynamics}.
\newblock Irwin/McGraw-Hill c2000.., 2010.

\bibitem[Tenenbaum \& Griffiths(2003)Tenenbaum and
  Griffiths]{tenenbaum2003theory}
Joshua~B Tenenbaum and Thomas~L Griffiths.
\newblock Theory-based causal inference.
\newblock In \emph{Advances in neural information processing systems}, pp.\
  43--50, 2003.

\bibitem[Trani et~al.(2016)Trani, Ballard, Bakhshi, and
  Hovmand]{trani2016community}
Jean-Francois Trani, Ellis Ballard, Parul Bakhshi, and Peter Hovmand.
\newblock Community based system dynamic as an approach for understanding and
  acting on messy problems: a case study for global mental health intervention
  in afghanistan.
\newblock \emph{Conflict and health}, 10\penalty0 (1):\penalty0 25, 2016.

\end{thebibliography}
\bibliographystyle{iclr2020_conference}

\end{document}